\documentstyle[11pt,epsfig,amssymb]{article}
\title{\bf Correspondence between modified gravity and $5D$ Ricci-flat cosmologies}
\author{K. Atazadeh$^{1}$\thanks{email: k-atazadeh@sbu.ac.ir},
\,\,\,F. Darabi$^{2,3}$\thanks{email: f.darabi@azaruniv.edu}\, and
H. R. Sepangi$^{1}$\thanks{email: hr-sepangi@sbu.ac.ir}
\\ $^{1}${\small Department of Physics, Shahid Beheshti University, Evin,
Tehran 19839, Iran}\\$^2${\small  Research Institute for Astronomy
and Astrophysics of Maragha (RIAAM), Maragha 55134-441,
Iran}\\$^3${\small Department of Physics, Azarbaijan University of
Tarbiat Moallem, Tabriz 53741-161, Iran}}
\begin{document}
\maketitle

\begin{abstract}
We study the correspondence between two theoretical frameworks for
describing dark energy, $f(R)$ gravity and higher-dimensional
Space-Time-Matter (STM) or induced-matter theory. We show that the
Hubble expansion parameter in $f(R)$ gravity can be associated with
a combination of metric functions in STM theory, and consider a
specific example whose properties are consistent with late-time
acceleration.
\end{abstract}
\vspace{2cm}

\section{Introduction}
The question of dark energy and the accelerating universe has been
the focus of a large amount of activities in recent years. This
expansion has directly been measured from the light-curves of
several hundred type Ia supernovae \cite{1,2} and independently from
observations of the cosmic microwave background (CMB) by the WMAP
satellite \cite{3} and other CMB experiments \cite{4,5}. All these
results strongly suggest that the universe is permeated smoothly by
dark energy which has a negative pressure and violates the strong
energy condition. Dark energy and the accelerating universe have
been discussed extensively from various point of views over the past
few years \cite{Quintessence,Phantom,K-essence}. In principle, a
natural candidate for dark energy could be a small cosmological
constant. One approach in this direction is to employ what is known
as modified gravity where an arbitrary function of the Ricci scalar
is added to the Einstein-Hilbert action. It has been shown that such
a modification may account for the late time acceleration and the
initial inflationary period in the evolution of the universe
\cite{modifidgravity2}. One of the first proposals in this regard
was suggested in \cite{modifidgravity} where a term of the form
$R^{-1}$ was added to the usual Einstein-Hilbert action. However,
several models of $f(R)$ gravity, proposed in
\cite{modifidgravity2,modifidgravity}, have been shown to suffer
from unwanted instabilities \cite{dolgov} or violate  solar system
constraints \cite{solar}. These instabilities arise because of an
extra propagating scalar degree of freedom, and hence any viable
$f(R)$ model must be constructed in such a way as to render the
dynamics of this scalar field carefully controllable. This is in
principle possible, and indeed $f(R)$ theories can be made to
satisfy solar system and laboratory tests by invoking the so-called
Chameleon mechanism \cite{chameleon}. Recently, the authors in
\cite{maeda} have tried to resolve the singularity problem arising
in the strong gravity regime of the otherwise viable $f(R)$ theories
by imposing appropriate fine-tuning on their models.

In $f(R)$ gravity, Einstein equations posses extra terms induced
from geometry which, when moved to the right hand side, may be
interpreted as a matter source represented by the energy-momentum
tensor $T^{\rm Curv}$, see equation (\ref{eq03}). In a similar
fashion, the Space-Time-Matter (STM) theory, discussed below,
results in Einstein equations in $4D$ with some extra geometrical
terms which may be interpret as induced matter. It therefore seems
plausible to make a correspondence between the geometrical terms in
STM and $T^{\rm Curv}$ resulting in $f(R)$ gravity. We shall explore
this idea to show that different choices of the parameter $\mu(t)$
in STM may correspond to different choices of $f(R)$ in curvature
quintessence models in modified gravity.

The correspondence discussed above is based on the idea of extra
dimensions. The idea that our world may have more than four
dimensions is due to Kaluza \cite{Kaluza}, who unified Einstein's
theory of General Relativity with Maxwell's theory of
Electromagnetism in a $5D$ manifold. Since then, higher dimensional
or Kaluza-Klein theories of gravity have been studied extensively
\cite{Klein} from different angles. Notable amongst them is the STM
theory mentioned above, proposed by Wesson and his collaborators,
which is designed to explain the origin of matter in terms of the
geometry of the bulk space in which our $4D$ world is embedded, for
reviews see \cite{Wesson}. More precisely, in STM theory, our world
is a hypersurface embedded in a five-dimensional Ricci-flat
($R_{AB}=0$) manifold where all the matter in our world can be
thought of as being  manifestations of the geometrical properties of
the higher dimensional space. The fact that such an embedding can be
done is supported by Campbell's theorem \cite{Compbell} which states
that any analytical solution of the Einstein field equations in $N$
dimensions can be locally embedded in a Ricci-flat manifold in
$\left(N+1\right)$ dimensions. Since the matter is induced from the
extra dimension, this theory is also called the induced matter
theory. Applications of the idea of induced matter or induced
geometry can also be found in other situations \cite{FSS}. The STM
theory allows for the metric components to be dependent on the extra
dimension and does not require the extra dimension to be compact.
The sort of cosmologies stemming from STM theory is studied in
\cite{LiuW,STM-cosmology,WLX}.

In this paper we consider the correspondence between $f(R)$ gravity
and STM theory. In section 2 we present a short review of $4D$ dark
energy models in the framework of $f(R)$ gravity. In section 3 the
field equations are solved in STM theory by fixing a suitable metric
and the resulting geometric terms are interpreted as dark energy.
The cosmological evolution in STM are considered in section 4.
Section 5 deals with an example for a special form of $f(R)$.
Conclusions are drawn in the last section.

\section{$4D$ dark energy models in modified gravity}
To begin with, let us start by a brief review of $f(R)$ gravity
theory. We start by writing the four dimensional action as
\begin{equation}\label{eq01}
{\cal S}=\int d^4x \sqrt{-g} f(R)+{\cal S}_{m},
\end{equation}
where $f(R)$ is a function of the Ricci scalar $R$ and ${\cal
S}_{m}$ is the action for the matter fields. We use units such
that $8\pi G_N=c=\hbar=1$. The field equations are
\begin{equation}\label{eq02}
f'(R)R_{\mu\nu}-\frac{1}{2}f(R)g_{\mu\nu}=
f'(R)^{;\alpha\beta}(g_{\alpha\mu}g_{\beta\nu}-g_{\alpha\beta}g_{\mu\nu})+
\tilde{T}^{m}_{\mu\nu}\,,
\end{equation}
which can be recast into the more expressive form
\begin{equation}\label{eq03}
G_{\mu\nu}=R_{\mu\nu}-\frac{1}{2}g_{\mu\nu}R=T^{\rm
Curv}_{\mu\nu}+T^{m}_{\mu\nu}\,,
\end{equation}
where an stress-energy tensor has been defined for the curvature
contribution
\begin{equation}\label{eq04}
 T^{\rm^{Curv}}_{\mu\nu}=\frac{1}{f'(R)}\left\{\frac{1}{2}g_{\mu\nu}\left[f(R)-Rf'(R)\right]+
f'(R)^{;\alpha\beta}(g_{\alpha\mu}g_{\beta\nu}-g_{\alpha\beta}g_{\mu\nu})
\right\},
\end{equation}
and
\begin{equation}\label{eq05}
 T^{m}_{\mu\nu}=\frac{1}{f'(R)}\tilde{T}^{m}_{\mu\nu}\,,
\end{equation}
is the stress-energy tensor of the matter. To derive the field
equations we consider the Robertson-Walker metric for the evolution
of the cosmos
\begin{equation}\label{eq06}
ds^{2}=dt^{2}-a(t)^{2}\left(\frac{dr^{2}}{1-kr^{2}}+r^{2}d\Omega^{2}\right),
\end{equation}
where $k$ is the curvature of the space, namely, $k=0, 1, -1$ for
the flat, closed and open universes respectively. Substituting the
above metric in equation (\ref{eq03}) we obtain the $4D$,
spatially flat Friedmann equations as follows
\begin{equation}\label{eq07}
H^2=\frac{1}{3}\left(\rho_{_{m}}+\rho_{_{\rm Curv}}\right),
\end{equation}
and
\begin{equation}\label{eq08}
\dot{H}=-\frac{1}{2}\left[(\rho_{_{m}}+p_{_{m}})+\rho_{_{\rm
Curv}}+p_{_{\rm Curv}}\right],
\end{equation}
where a dot represents derivation with respect to time. Such a
universe is dominated by a barotropic perfect fluid with the
equation of state (EOS) given by $ p_{_{m}}=w_{_{m}}\rho_{_{m}}$
($w_{m}=0$ for pressureless cold dark matter and $w_{m}=1/3$ for
radiation) and a spatially homogenous curvature quintessence.

The energy density and pressure of the curvature quintessence are
\begin{equation}\label{eq09}
p_{_{\rm
Curv}}=\frac{1}{f'(R)}\left\{2\left(\frac{\dot{a}}{a}\right)\dot{R}f''(R)+\ddot{R}f''(R)+\dot{R}^2f'''(R)
-\frac{1}{2}\left[f(R)-Rf'(R)\right] \right\},
\end{equation}
and
\begin{equation}\label{eq10}
\rho_{_{\rm
Curv}}=\frac{1}{f'(R)}\left\{\frac{1}{2}\left[f(R)-Rf'(R)\right]
-3\left(\frac{\dot{a}}{a}\right)\dot{R}f''(R) \right\},
\end{equation}
respectively. The equation of state of the curvature quintessence is
\begin{equation}\label{eq11}
w_{_{\rm Curv}}=\frac{p_{_{\rm Curv}}}{\rho_{\rm
_{Curv}}}.\label{eosscalar}
\end{equation}
Recently, cosmological observations  have indicate that our universe
is undergoing an accelerated expanding phase. This could be due to
the vacuum energy or dark energy which dominates our universe
against other forms of matter such as dark matter and Baryonic
matter. We thus concentrate on the vacuum sector {\it i.e.}
$\rho_{m}=p_{m}=0$, from which the evolution equation of curvature
quintessence becomes
\begin{equation}\label{eq12}
\dot{\rho}_{_{\rm Curv}}+3H\left(\rho_{_{\rm Curv}}+p_{_{\rm
Curv}}\right)=0,
\end{equation}
yielding
\begin{eqnarray}\label{eq13}
\rho_{_{\rm Curv}}(z)&=&\rho_{0_{\rm Curv}
}\exp\left[3\int_{0}^{z}(1+w_{_{\rm Curv}})d\ln(1+z)\right]\nonumber\\
&\equiv&\rho_{0_{\rm Curv} }E(z),
\end{eqnarray}
where, $1+z=\frac{a_0}{a}$ is the redshift and the subscript $0$
denotes the current value.  In terms of the redshift, the first
Friedmann equation can be written as
\begin{equation}\label{eq14}
H(z)^2=H_0^2\Omega_{0_{\rm Curv}}E(z),\label{Hz}
\end{equation}
where $\Omega_{0_{\rm Curv}}$ and $H_0$ are the current values of
the dimensionless density parameter and Hubble parameter,
respectively. Equation (\ref{eq14}) is the Friedmann equation in
terms of the redshift, $z$, which is suitable for cosmological
observations. In fact, equations (\ref{eq14}) and (\ref{eq26}),
obtained in section 4, are the cosmological connections between
$f(R)$ gravity  and STM theory.

\section{Dark energy in $5D$ models}\label{DE5}

In the context of STM theory, a class of exact $5D$ cosmological
solutions has been investigated and discussed in \cite{LiuM}. This
solution was further pursued in \cite{LiuW} where it was shown to
describe a cosmological model with a big bounce as opposed to the
ubiquitous big bang. The $5D$ metric of this solution reads
\begin{equation}\label{eq15}
dS^{2}=B^{2}dt^{2}-A^{2}\left(
\frac{dr^{2}}{1-kr^{2}}+r^{2}d\Omega ^{2}\right) -dy^{2},
\label{5-metric}
\end{equation}
where $d\Omega ^{2}\equiv \left( d\theta ^{2}+\sin ^{2}\theta d\phi
^{2}\right)$ and
\begin{eqnarray}\label{eq16}
A^{2}&=&\left( \mu ^{2}+k\right) y^{2}+2\nu y+\frac{\nu ^{2}+K}{\mu
^{2}+k},
\nonumber \\
B&=&\frac{1}{\mu }\frac{\partial A}{\partial t}\equiv
\frac{\dot{A}}{\mu }.
\end{eqnarray}
Here $\mu =\mu (t)$ and $\nu =\nu (t)$ are two arbitrary functions
of $t$, $k$ is the $3D$ curvature index $\left(k=\pm 1,0\right)$,
and $K$ is a constant. This solution satisfies the $5D$ vacuum
equation $R_{AB}=0$. The Kretschmann curvature scalar
\begin{eqnarray}\label{eq17}
I_{3} &=&R_{ABCD}R^{ABCD}=\frac{72K^{2}}{A^{8}},
\end{eqnarray}
shows that $K$ determines the curvature of the $5D$ manifold. Such a
solution was considered in \cite{LiuM} with a different notation.

Using the $4D$ part of the $5D$ metric (\ref{eq15}) to calculate
the $4D$ Einstein tensor, we obtain
\begin{eqnarray}\label{eq18}
^{(4)}G_{0}\,^{0} &=&\frac{3\left( \mu ^{2}+k\right) }{A^{2}},
\nonumber \\
^{(4)}G_{1}\,^{1} &=&^{(4)}G_{2}\,^{2}=^{(4)}G_{3}\,^{3}=\frac{2\mu \dot{\mu}}{A%
\dot{A}}+\frac{\mu ^{2}+k}{A^{2}}.
\end{eqnarray}
As was mentioned earlier, since the recent observations show that
the universe is currently going through an accelerated expanding
phase, we assume that the induced matter contains only dark energy
with $\rho_{_{DE}}$, {\it i.e.} $\rho_{_{m}}=0$. We then have
\begin{equation}\label{eq19}
\frac{3\left( \mu ^{2}+k\right) }{A^{2}}=\rho_{_{DE}},
\end{equation}
\begin{equation}\label{eq20}
\frac{2\mu \dot{\mu}}{A\dot{A}}+\frac{\mu
^{2}+k}{A^{2}}=-p_{_{DE}}.
\end{equation}
From equations (\ref{eq19}) and (\ref{eq20}), one obtains the EOS
of dark energy
\begin{equation}\label{eq21}
w_{_{DE}}=\frac{p_{_{DE}}}{\rho_{_{DE}}}=-\frac{2\left. \mu
\dot{\mu}\right/ A \dot{A}+\left. \left( \mu ^{2}+k\right) \right/
A^{2}}{3\left. \left( \mu ^{2}+k\right) \right/ A^{2}}.
\end{equation}
The Hubble and deceleration parameters are given in
\cite{LiuW,WLX} and can be written as
\begin{eqnarray}\label{eq22}
H&\equiv&\frac{\dot{A}}{A B}=\frac{\mu}{A},
\end{eqnarray}
and
\begin{eqnarray}\label{eq23}
q \left(t, y\right)&\equiv&\left.
-A\frac{d^{2}A}{dt^{2}}\right/\left(\frac{dA}{dt}\right)^{2}
=-\frac{A \dot{\mu}}{\mu \dot{A}},
\end{eqnarray}
from which we see that $\dot{\mu}\left/\mu\right.>0$ represents an
accelerating universe while $\dot{\mu}\left/\mu\right.<0$ represents
a decelerating one. The function $\mu(t)$ therefore plays a crucial
role in defining the properties of the universe at late times.

\section{ Cosmological evolution of STM theory}
In this section we will concentrate on the predictions of the
cosmological evolution in the spatially flat case ($k=0$). To
avoid having to specify the form of the function $\nu(t)$, we
change the parameter $t$ to $z$ and use $A_{0}\left/A \right.=1+z$
and define $\mu_{0}^{2}\left/ \mu^{2}\right.=F\left(z\right)$,
noting that $F(0)\equiv 1$. We then find that equations
(\ref{eq21})-(\ref{eq23}) reduce to
\begin{eqnarray}\label{eq24}
w_{_{DE}}(z) &=&-\frac{1+\left(1+z\right)d\ln
F\left(z\right)\left/dz\right.}{3},
\end{eqnarray}
and
\begin{eqnarray}\label{eq25}
q_{_{DE}}(z)&=&\frac{1+3\Omega_{_{DE}}w_{_{DE}}}{2}=-\frac{\left(1+z\right)}{2}\frac{d\ln
F\left(z\right)}{dz}.
\end{eqnarray}
There is an arbitrary function $\mu(t)$ in the present $5D$ model.
Different choices of $\mu(t)$ may correspond to different choices of
$f(R)$ in curvature quintessence models in modified gravity. Various
choices of $\mu(t)$ correspond to the choices of $F(z)$. This
enables us to look for the desired properties of the universe via
equations (\ref{eq24}) and (\ref{eq25}). Using these definitions,
the Friedmann equation becomes
\begin{equation}\label{eq26}
H^2=H_0^2(1+z)^2F(z)^{-1}. \label{friedmann}
\end{equation}
This would allow us to use the supernovae observational data to
constrain the parameters contained in the model or the function
$F(z)$. By comparing equation (\ref{eq26}) with equation
(\ref{eq14}), we find that there exists a correspondence between the
functions $f(R)$ and $F(z)$. We thus take $F(z)$ as
\begin{equation}\label{eq27}
F(z)=(1+z)^2\left[\Omega_{0_{\rm Curv}}E(z)\right]^{-1}.\label{fz}
\end{equation}
According to (\ref{eq13}), it is easy to see that the function
$E(z)$ is determined by the particular choice of $f(R)$ which, in
turn, determines the function $F(z)$ through equation (\ref{eq27}).
The evolution of the density components and EOS of dark energy may
now be derived. In the next section we will consider the
correspondence between modified gravity and the Ricci-flat cosmology
for a generic $f(R)$.

\section{Example}
To continue, we must determine the functional form of $f(R)$. Thus,
for example, we choose $f(R)$ as a generic power law of the scalar
curvature and assume for the scale factor a power law solution in
$4D$, investigated in the first reference in \cite{modifidgravity2}.
Therefore
\begin{equation}\label{eq28}
f(R)=f_0 R^n\,,\qquad
a(t)=a_0\left(\frac{t}{t_0}\right)^{\alpha}\,.
\end{equation}
The interesting cases are for the values of $\alpha$ satisfying
$\alpha> 1$ which would lead to an accelerated expansion of our
universe. Let us now concentrate on the case $\rho_{_{m}}=0$.
Inserting equation (\ref{eq28}) into the dynamical system
(\ref{eq07}) and (\ref{eq08}), for a spatially flat space-time we
obtain an algebraic system for parameters $n$ and $\alpha$
 \begin{equation} \label{eq29}
 \left\{ \begin{array}{ll} \alpha \left[\alpha(n-2)+2n^{2}-3n+1\right]=0, \\
 \\
\alpha\left[n^{2}-n+1+\alpha(n-2)\right]=n(n-1)(2n-1),\\
\end{array}
\right.
\end{equation}

\vspace{7mm}

\noindent from which the allowed solutions are
\begin{equation}\label{eq30}
\begin{array}{cc} \alpha=0\,\, \rightarrow\,\, n=0,\,\,1/2,\,\,1,\\ \\
\alpha=\displaystyle\frac{2n^2-3n+1}{2-n}\,,\,\, \forall{n},\,\
\,  n\neq {2}\,.
\end{array}
\end{equation}
\vspace{7mm} The solutions with $\alpha=0$ are not interesting since
they provide static cosmologies with a non-evolving scale factor.
Note that this result matches the standard General Relativity
result, $n=1$, in the absence of matter. On the other hand, the
cases with generic $\alpha$ and $n$ furnish an entire family of
significant cosmological models. Using equations (\ref{eq09}) and
(\ref{eq10}) we can also deduce the equation of state for the family
of solutions $\alpha=\displaystyle\frac{2n^2-3n+1}{2-n}$ as
\vspace{7mm}
\begin{equation}\label{eq31}
w_{_{\rm Curv}}(n)=-\left(\frac{6n^2-7n-1}{6n^2-9n+3}\right)\,,
\end{equation}
where $w_{_{\rm Curv}}\rightarrow{-1}$ as $n\rightarrow {\infty}$.
This shows that an {\it infinite} $n$ is compatible with recovering
an {\it infinite} cosmological constant. Thus, using equation
(\ref{eq31}), the functions $E(z)$ and $F(z)$ are given by
\begin{eqnarray}\label{eq32}
\label{pot1} E(z)&=&(1+z)^{3\left[\frac{-2n+4}{6n^2-9n+3}\right]},
\end{eqnarray}
\begin{eqnarray}\label{eq33}
\label{pot1} F(z)&=&(1+z)^2\left[\Omega_{0_{\rm Curv}
}(1+z)^{3\left[\frac{-2n+4}{6n^2-9n+3}\right]}\right]^{-1}.
\end{eqnarray}
Now, using the above equations, we see that equations (\ref{eq24})
and (\ref{eq25}) can be written as
\begin{equation}\label{eq34}
w_{_{DE}}(n)=-\left(\frac{6n^2-7n-1}{6n^2-9n+3}\right),
\end{equation}
and
\begin{equation}\label{eq35}
q_{_{DE}}(n)\equiv-\frac{A\dot{\mu}}{\mu\dot{A}}=\frac{-2n^{2}+2n+1}{2n^{2}-3n+1}.
\end{equation}
Therefore, within the context of the present investigation, the
accelerating, dark energy dominated universe can be obtained by
using the correspondence between $F(z)$ and $f(R)$ in modified
gravity theories. We observe that in STM theory, $5D$ dark energy
cosmological models correspond to $4D$ curvature quintessence
models. This result is consistent with the correspondence between
exact solutions in Kaluza-Klein gravity and scalar tensor theory
\cite{billyard}. We also note that, as is well known, with a
suitable conformal transformation, $f(R)$ gravity reduces to scalar
tensor theory.

From equations (\ref{eq32}) and (\ref{eq33}), we can rewrite
equation (\ref{eq26}) as
\begin{equation}\label{eq36}
h(z, n)=\Omega_{0_{\rm Curv}}(1+z)^{3\left[\frac{-2n+4}{6n^2-9n+3}\right]},
\end{equation}
where $h(z,n)\equiv\frac{H(z)^{2}}{H_0^{2}}$ and the contribution of
ordinary matter has been neglected. Figure 1 shows the behavior of
$h(n)$ as a function of $n$ for $z\sim1.5$ and $\Omega_{0_{\rm
Curv}}\simeq0.70$. As can be seen, for $n\longrightarrow\pm\infty$
and $z\longrightarrow0$ we have $h(z,n)\longrightarrow\Omega_{0_{\rm
Curv}}$, that is, the universe finally approaches the curvature
dominant state, thus undergoing an accelerated expanding phase.
Figure 2 shows the behavior of $h(z)$ as a function of $z$ for
$n=2,10,-10$ and $\Omega_{0_{\rm Curv}}\simeq0.70$. We see that for
small $z$, $h(z)\longrightarrow0.70$. Thus, we have obtained
late-time accelerating solutions only by using the correspondence
between $f(R)$ gravity and STM theory. Here, we have interpreted the
properties of $5D$ Ricci-flat cosmologies by dark energy models in
modified gravity.
\begin{figure}
\begin{center}
\epsfig{figure=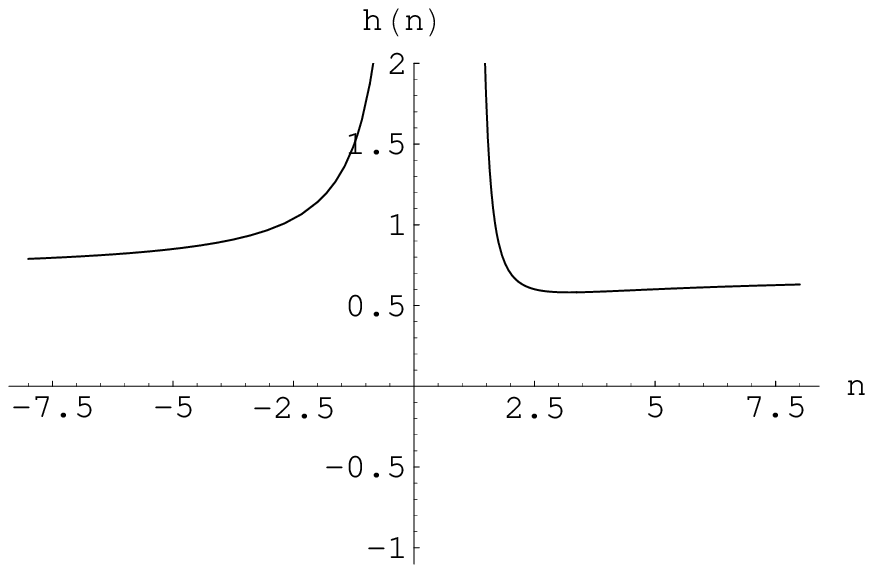,width=9cm}\hspace{5mm}
\end{center}
\caption{\footnotesize Behavior of $h(n)$ as a function
of $n$ for $z\sim1.5$ and
$\Omega_{0_{\rm Curv}}\simeq0.70$. An accelerating universe occurs for
$n\lesssim-2$ and $n \gtrsim 2$.}
\begin{center}
\epsfig{figure=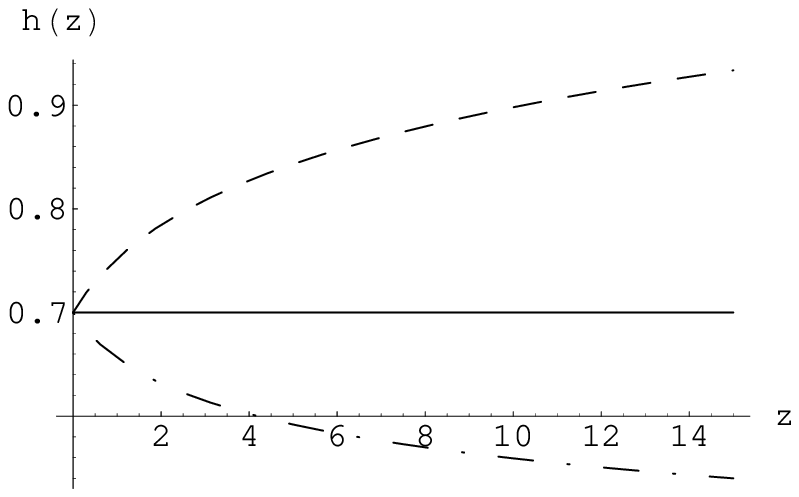,width=7cm}\hspace{5mm}
\end{center}
\caption{\footnotesize Behavior of $h(z)$ as a function of $z$ for
$n=2$ (solid line), $n=10$ (dashed line), $n=-10$ (dot-dashed line)
and $\Omega_{0_{\rm Curv}}\simeq0.70$. It can be seen that for the
above values of $n$, as $z\rightarrow0$, the curves converge to the
point $h(z)\rightarrow0.7$.}
\end{figure}

\section{Conclusions}
In this paper we have investigated the present accelerated expanding
phase of the universe using a general class of $5D$ Ricci-flat
cosmological models. Such an exact solution contains two arbitrary
functions, $\mu(t)$ and $\nu(t)$  which are analogous to different
forms of $f(R)$ in curvature quintessence models. Also, once the
forms of the arbitrary functions are specified, the characteristic
parameters determining the evolution of our universe are specified.
We have noted that the correspondence between the functions $F(z)$
and $f(R)$ plays a crucial role and defines the form of the function
$F(z)$. Finally, by taking a specific form for $f(R)$ we obtained
solutions that are consistent with late-time acceleration of the
universe.
\section*{Acknowledgment}

This work has been financially supported  by the Research
Institute for Astronomy and Astrophysics of Maragha.


\begin{thebibliography}{9}
\bibitem{1}A. G. Riess {\it et. al.}  [High-Redshift Supernova Search Team, (HZT) ], Astron. J. 116
(1998) 1006, [astro-ph/9805201].
\bibitem{2}S. Perlmutter {\it et. al.}, Astron. J. 517 (1999) 565, [astro-ph/9812133];\\
D. N. Spergel {\it et. al.}, Astrophys. J. Suppl. 148 (2003) 175,
[astro-ph/0302209].
\bibitem{3}C. L. Bennett {\it et. al.}, Astrophys. J. Suppl. 148 (2003) 1, [astro-ph/0302207].
\bibitem{4}C. B. Netterfield{\it et. al.}, Astrophys. J. 571  (2002) 604,
[astro-ph/0104460].
\bibitem{5}N. W. Halverson {\it et. al.}, Astrophys. J.  568 (2002)  38, [astro-ph/0104489].
\bibitem{Quintessence} I. Zlatev, L. Wang and P. J. Steinhardt ,
Phys. Rev. Lett. 82 (1999) 896, [astro-ph/9807002];\\
P. J. Steinhardt, L. Wang, I. Zlatev, Phys. Rev. D 59 (1999) 123504,
[astro-ph/9812313];\\
M. S. Turner, Int. J. Mod. Phys. A  17S1 (2002) 180,
[astro-ph/0202008];\\
V. Sahni, Class. Quant. Grav. 19 (2002) 3435, [astro-ph/0202076].
\bibitem{Phantom} R. R. Caldwell, M. Kamionkowski,
N. N. Weinberg, Phys. Rev. Lett. 91 (2003) 071301,
[astro-ph/0302506];\\
R. R. Caldwell, Phys. Lett. B 545 (2002) 23, [astro-ph/9908168];\\
P. Singh, M. Sami, N. Dadhich, Phys. Rev. D 68 (2003) 023522,
[hep-th/0305110];\\
J.G. Hao, X.Z. Li, Phys. Rev. D 67 (2003) 107303, [gr-qc/0302100].
\bibitem{K-essence} Armend\'{a}riz-Pic\'{o}n, T. Damour, V. Mukhanov,
Phy. Lett. B  458 (1999) 209 ;\\
M. Malquarti {\it et. al.},  Phys. Rev. D
67 (2003) 123503;\\
T. Chiba , Phys. Rev. D 66 (2002) 063514, [astro-ph/0206298].

\bibitem{modifidgravity2}
S. Capozziello, Int. J. Mod. Phys. D 11 (2002) 483,
[astro-ph/0201033];\\ S. Capozziello {\it et. al.}, Int. Mod. Phys.
D 12 (2003) 1969, [astro-ph/0307018];\\S. Capozziello {\it et. al.},
Phys. Lett.
B 639 (2006) 135;\\
S. Nojiri and S. D. Odintsov,  J. Phys. Conf. Ser. 66 (2007)
012005;\\ S. Nojiri and S. D. Odintsov, J. Phys. A 40 (2007) 6725;\\
F. Briscese {\it et. al.}, Phys. Lett.
B 646 (2007) 105;\\
S. Nojiri, S.D. Odintsov and H. Stefan\'{c}ic, Phys. Rev. D 74,
(2006) 086009;\\
M. E. Soussa and R. P. Woodard, Gen. Rel. Grav. 36 (2004) 855;\\
R. Dick, Gen. Rel. Grav. 36 (2004) 217;\\
A. E. Dominguez and D. E. Barraco,  Phys. Rev. D 70, (2004) 043505;\\
V. Faraoni, Phys. Rev. D 75 (2007) 067302;\\
J. C. C. de Souza and V. Faraoni, Class. Quant. Grav. 24 (2007)
3637;\\
D. A. Easson,  Int. J. Mod. Phys. A 19 (2004) 5343;\\
G. J. Olmo, Phys. Rev. Lett. 95 (2005) 261102;\\
G. Allemandi {\it et. al.}, Gen.
Rel. Grav. 37 (2005) 1891;\\
S. Capozziello and A. Troisi, Phys. Rev. D 72 (2005) 044022 ;\\
T. Clifton and J.D. Barrow, Phys. Rev. D 72 (2005) 103005;\\
T. P. Sotiriou, Gen. Rel. Grav. 38 (2006) 1407;\\
S. Capozziello, A. Stabile and A. Troisi,  Mod. Phys. Lett. A 21
(2006) 2291;\\
A. Dolgov and D. N. Pelliccia,  Nucl. Phys. B 734 (2006) 208 ;\\
K. Atazadeh and H.R. Sepangi, Int. J. Mod. Phys. D 16 (2007) 687,
[gr-qc/0602028];\\
 S. Nojiri and S. D. Odintsov, Int. J. Geom. Meth. Mod. Phys. 4 (2007)
 115, [hep-th/ 0601213].

\bibitem{modifidgravity}
S. M. Carroll {\it et. al.}, Phys. Rev. D 70  (2004) 043528.
\bibitem{dolgov}A. D. Dolgov and M. Kawasaki, Phys. Lett. B 573 (2003)
1, [astro-ph/0307285].
\bibitem{solar}T. Chiba, Phys. Lett. B 575 (2003) 1,
[astro-ph/0307338];\\A. L. Erickcek, T. L. Smith and M.
Kamionkowski, Phys. Rev. D 74 (2006) 121501,  [astro-ph/0610483];\\
T. Chiba, T. L. Smith and A. L. Erickcek, Phys. Rev. D 75 (2007)
124014, [astro-ph/0611867].
\bibitem{chameleon} J. Khoury and A. Weltman, Phys. Rev. Lett. 93 (2004)
171104, [astro-ph/0309300];\\ J. Khoury and A. Weltman, Phys. Rev. D
69 (2004) 044026,  [astro-ph/0309411];\\ D. F. Mota and J. D.
Barrow, Mon. Not. Roy. Astron. Soc. 349 (2004) 291,
[astro-ph/0309273];\\ T. Clifton, D. F. Mota and J. D. Barrow,
Mon. Not. Roy. Astron. Soc. 358 (2005) 601,  [gr-qc/0406001];\\
 D. F. Mota and D. J. Shaw, Phys. Rev. Lett. 97 (2006) 151102,
[hep-ph/0606204];\\ D. F. Mota and D. J. Shaw, Phys. Rev. D 75
(2007) 063501,  [arXiv:hep-ph/0608078].

\bibitem{maeda}T. Kobayashi and K. Maeda, [arXiv:0810.5664].
\bibitem{Kaluza} T. Kaluza, {\it On The Problem Of Unity In Physics}, Sitzungsber. Preuss. Akad. Wiss.
Berlin Math. Phys. Kl  (1921) 966.
\bibitem{Klein} O. Klein, {\it Quantum Theory And Five-Dimensional Relativity}, Z. Phys. {\bf 37} 895 (1926)
[Surveys High Energy Phys. 5 (1986) 241].
\bibitem{Wesson} P. S. Wesson, \textit{Space-Time-Matter} (Singapore: World
Scientific) 1999;\\
J. M. Overduin and P. S. Wesson, Phys. Rept. 283 (1997) 303,
[gr-qc/9805018].
\bibitem{Compbell} J. E. Campbell, {\it A Course of Differential Geometry},
(Clarendon Oxford, 1926);\\
S. Rippl, R. Romero, R. Tavakol, Class. Quant. Grav. 12 (1995)
2411;\\
C. Romero and R. Tavakol and R. Zalaletdinov, Gen. Rel. Grav. 28
 (1996) 365;\\
 J. E. Lidsey {\it et. al.},
Class. Quant. Grav. 14 (1997) 865;\\
S. S. Seahra and P. S. Wesson, Class. Quant. Grav. 20 (2003) 1321,
[gr-qc/0302015].
\bibitem{FSS} V. Frolov, M. Snajdr and D. Stojkovic,  Phys. Rev. D 68  (2003) 044002.
\bibitem{LiuW} H. Y. Liu and P. S. Wesson, Astrophys. J.
562  (2001) 1, [gr-qc/0107093].
\bibitem{STM-cosmology} T. Liko and P. S. Wesson, Int. J. Mod. Phys. A 20 (2005) 2037,
[gr-qc/0310067];\\
 S. S. Seahra and P. S. Wesson,  J. Math. Phys.
44 (2003) 5664;\\
J. Ponce de Leon, Gen. Rel. Grav. 20  (1988) 539;\\
L. X. Xu, H. Y. Liu, B. L. Wang, Chin. Phys. Lett. 20 (2003) 995,
[gr-qc/0304049];\\
 H. Y. Liu,  Phys. Lett. B 560 (2003) 149,
[hep-th/0206198].
\bibitem{WLX} B. L. Wang, H. Y. Liu and L.X. Xu,  Mod. Phys. Lett. A  19  (2004) 449, [gr-qc/0304093];\\
L. X. Xu and H. Y. Liu, Int. J. Mod. Phys. D 14  (2005) 883,
[astro-ph/0412241].
\bibitem{LiuM} H. Y. Liu and B. Mashhoon, Ann. Phys. 4 (1995) 565.

\bibitem{billyard}A. Billyard and A. Coley, Mod. Phys. Lett. A 12
(1997) 2121.
\end{thebibliography}
\end{document}